\documentclass[pre,twocolumn,showpacs,amsmath,amssymb]{revtex4}

\usepackage{graphicx}% Include figure files
\usepackage{dcolumn}% Align table columns on decimal point
\usepackage{bm}% bold math

\begin{document}

\newcommand{\la}{\langle}
\newcommand{\ra}{\rangle}

%-------------------------fig.1------------------------------------

\newcommand{\figone}{%
  \begin{figure}
     \centering
     \begin{tabular}{cc}
        \includegraphics[height=3cm, width=4.3cm]{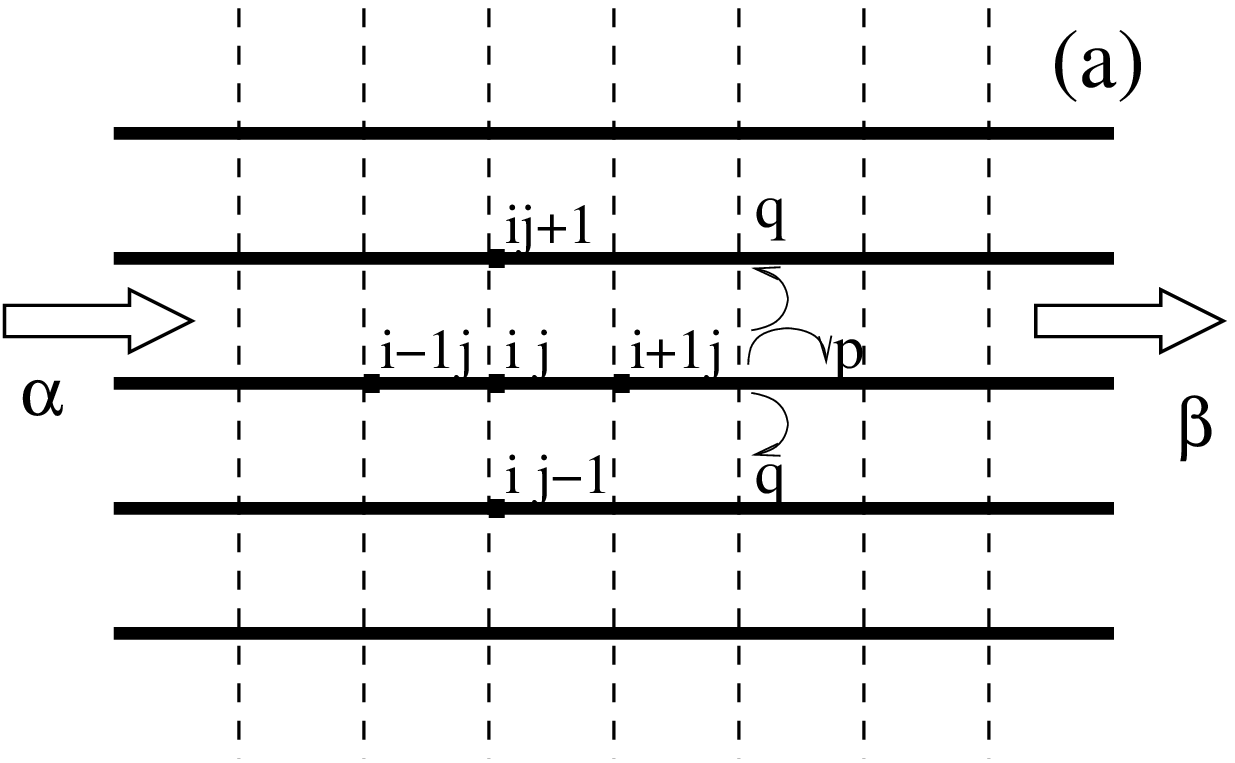}&
         \includegraphics[height=3cm, width=4.2cm]{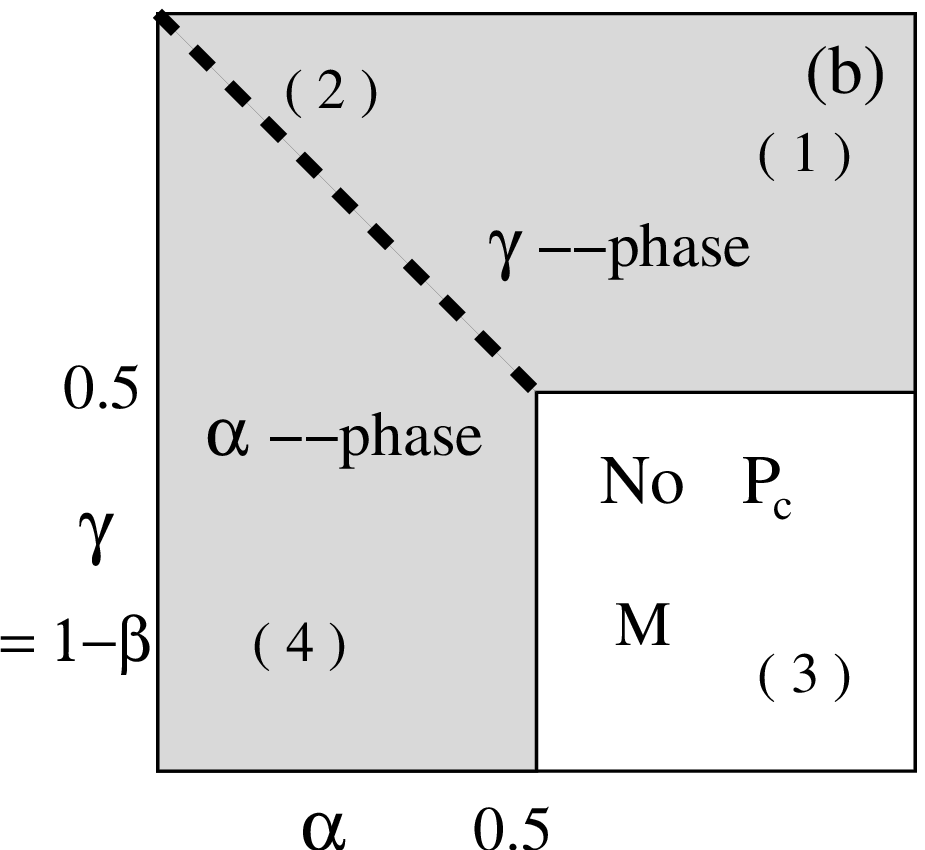}
     \end{tabular}
     \caption{%
       (a) Schematic diagram of 2-D asymmetric simple exclusion
       process (ASEP), with forward jump probability $p$ (thick lines)
       and transverse (dashed lines) excursion probability $q$, with
       $p+2q = 1$. For $q=0$, one gets decoupled 1-D ASEP.  (b)The
       phase diagram of 1-D and 2-D exclusion process. M $\equiv$
       maximal current phase ($\rho=0.5$), in the $\alpha$-phase the
       bulk density is $\alpha$, and in the $\gamma$-phase it is
       $\gamma$.  The dotted (thin solid) line is the first order
       (continuous) phase boundary.  For the 2-D or the modified 1-D,
       the M region widens with $p$ inducing a transition to M in the
       shaded region.  In the $\alpha$-phase region $p_c =2\alpha$ and
       in the $\gamma$-phase region $p_c = 2\beta$.   The points marked
       refer to Fig.  \ref{fig:3} }%
  \label{fig:1}
 \end{figure}
}%

%-------------------end fig 1-----------------------------------

%-------------------fig.2--------------------------

\newcommand{\figtwo}{%
   \begin{figure}
      \begin{tabular}{cc}
        \includegraphics[height=3.5cm, width=4.2cm]{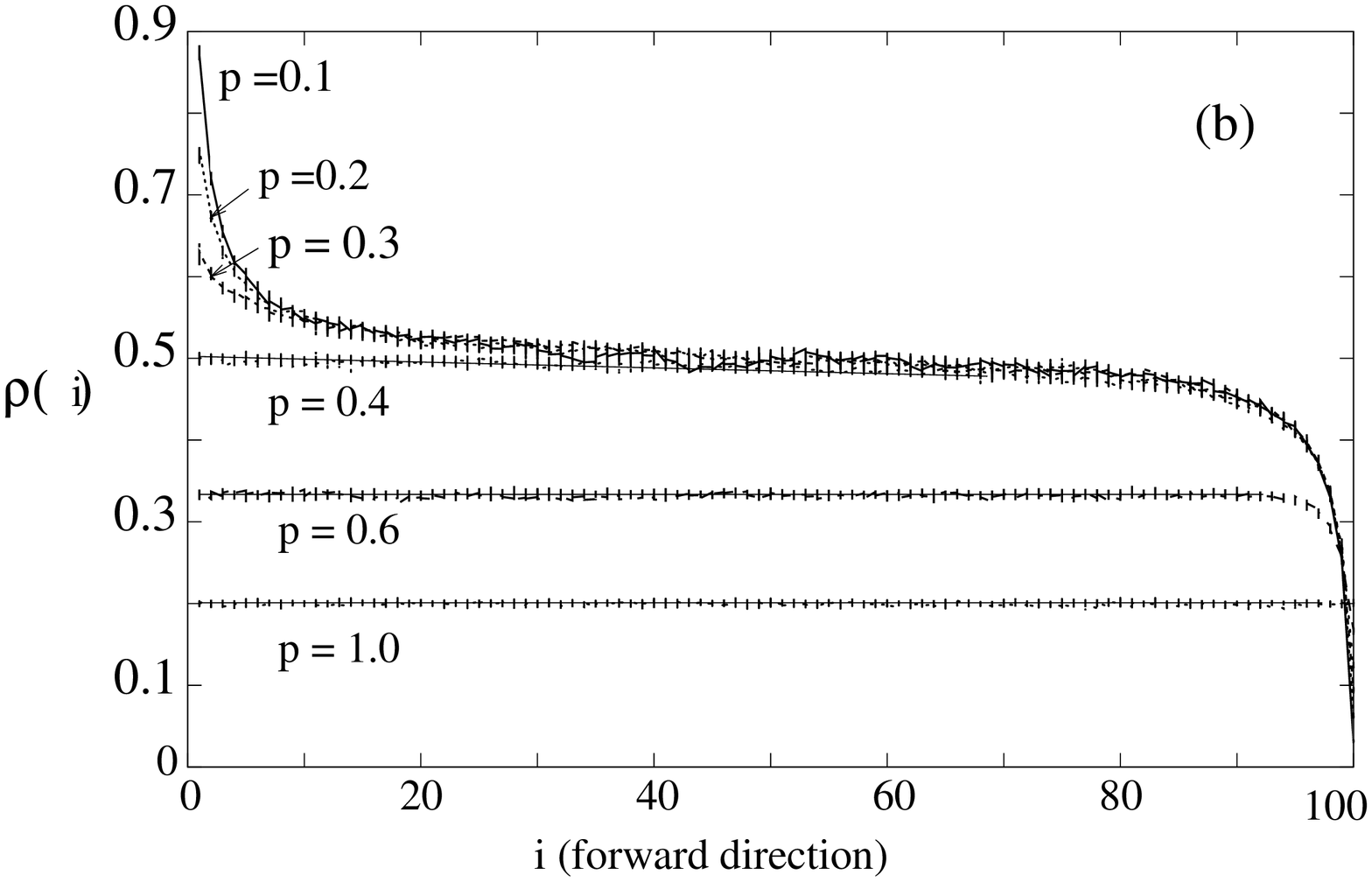}&
	\includegraphics[height=3.2cm, width=4.2cm]{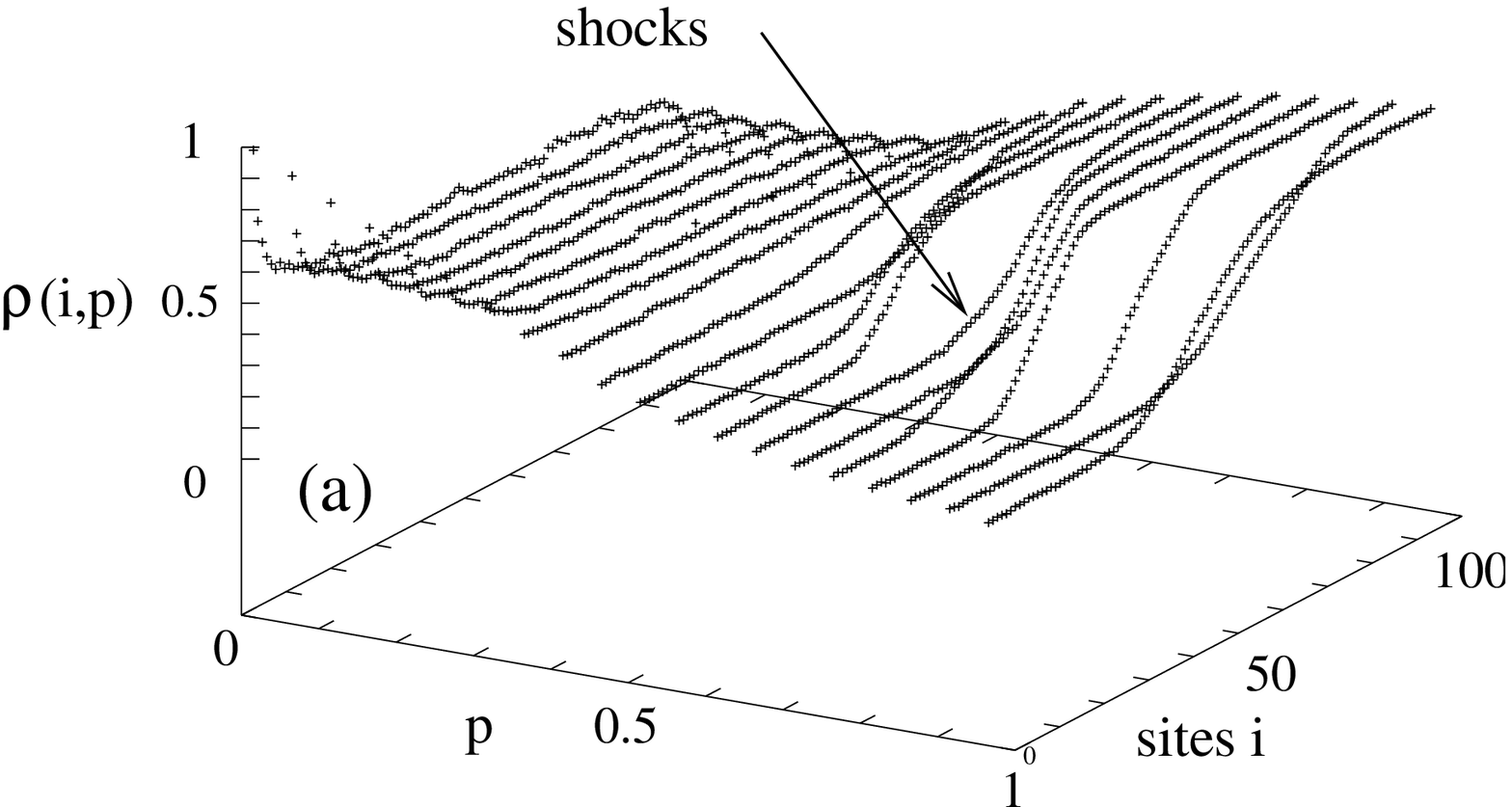}
      \end{tabular}
       \caption{(a):  
       Spatial density
       distribution $\rho(i,j)$ for a particular $j$ along the forward
       direction $i$ for various values of $p$ ($\alpha=\gamma=0.2$).
       The vertical width
       of any point on the line reflects the variation in the density
       in the transverse direction. For $p < p_c$ curves form one
       group and for $p > p_c$ the bulk density is determined by the
       left boundary  $\rho(1,j)=\alpha/p$. (b): The presence of diffusing 
       shocks for $\alpha =0.2$ and $\gamma = 0.8$ when $p > p_c$.}
      \label{fig:2}
    \end{figure}
}%
%-------------------end fig 2----------------------------------

%------------------fig 3 ----------------------------------

\newcommand{\figthree}{%
  \begin{figure}
    \centering
    \begin{tabular}{cc}
      \includegraphics[height=3.3cm, width=4.3cm]{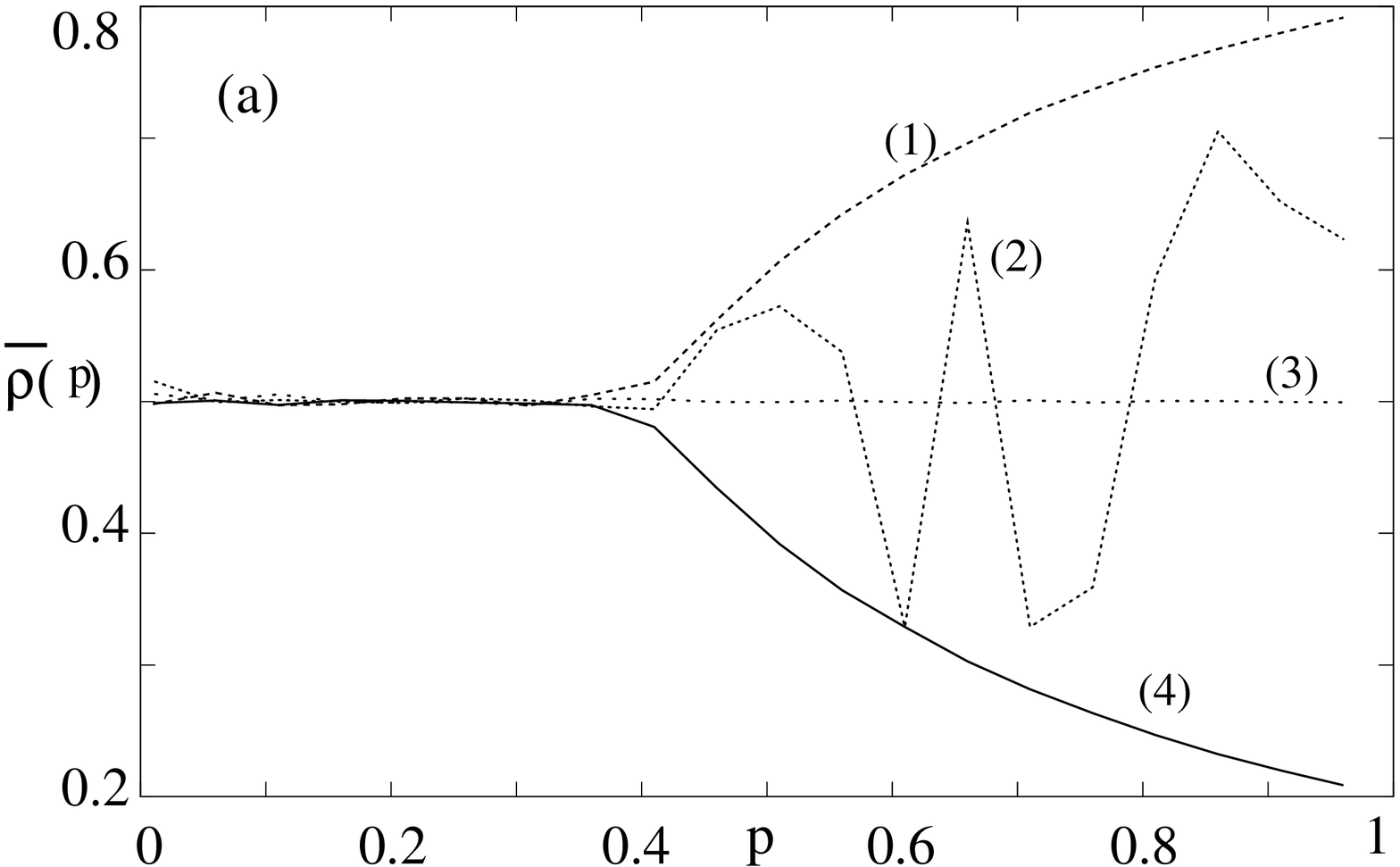}&
      \includegraphics[height=3.3cm, width=4.3cm]{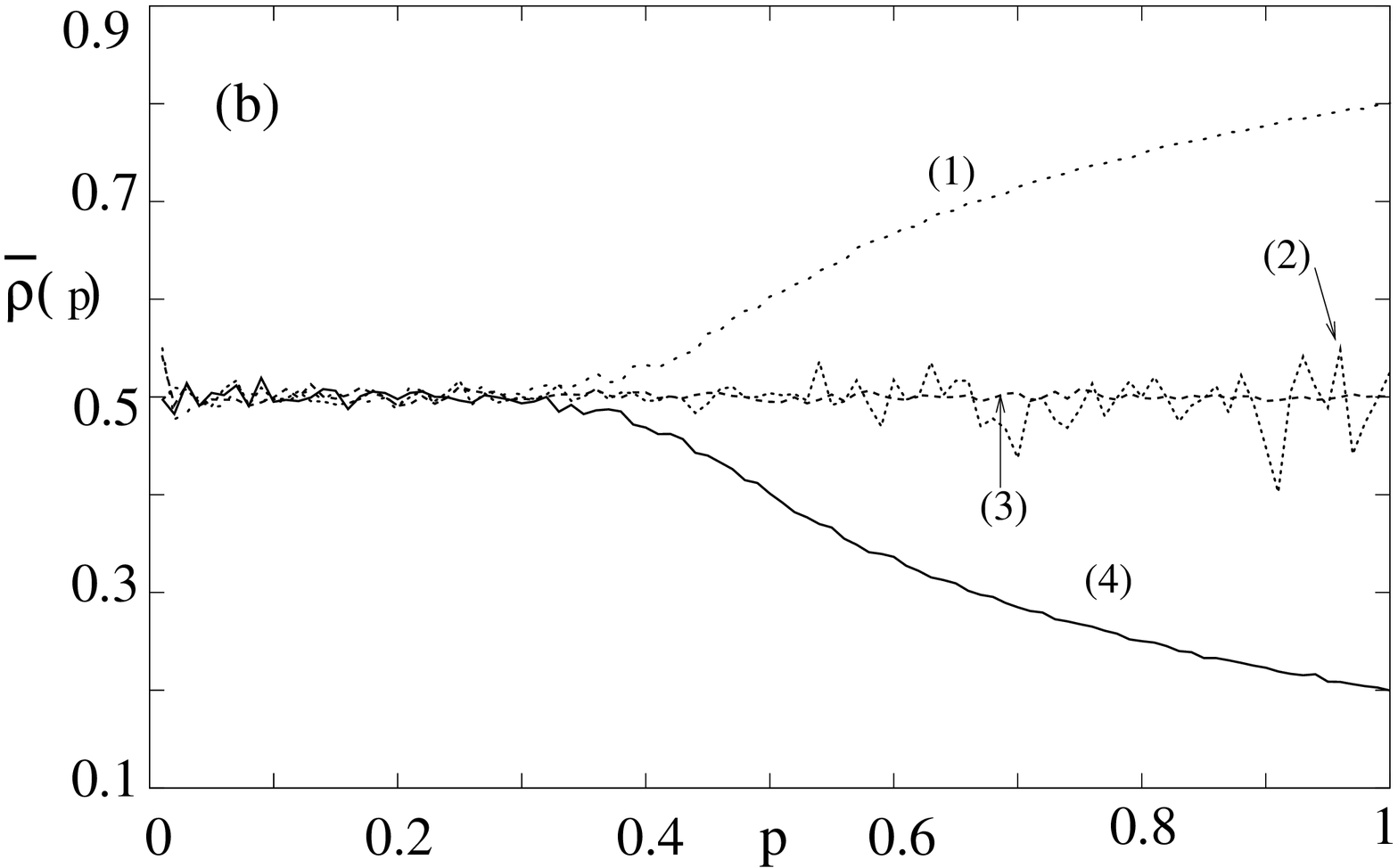}
    \end{tabular}
    \caption{The density phase transition for various values of
      $\alpha$ and $\beta$ in (a) 2-D and (b) 1-D cases. The average
      bulk density $\bar{\rho}=0.5$ up to $p < p_c$, but varies with
      $p$ for $p>p_c$. For both, curve (1) is for $\alpha = 0.8$,
      $\beta = 0.2$, curve (2) for $\alpha = 0.2$, $\beta = 0.2$,
      curve (3) for $\alpha = 0.8$, $\beta = 0.8$, and curve (4) is
      for $\alpha = 0.2$, $\beta = 0.8$, as marked in Fig.
      \ref{fig:1} (b).  All these have same $p_c$.  }
    \label{fig:3}
  \end{figure}
}%

%-------------------end fig 3-----------------------------------

%----------------------fig 4------------------------------

\newcommand{\figfour}{%
%\begin{widetext}
  \begin{figure}
    \centering
        \includegraphics[ width=6cm]{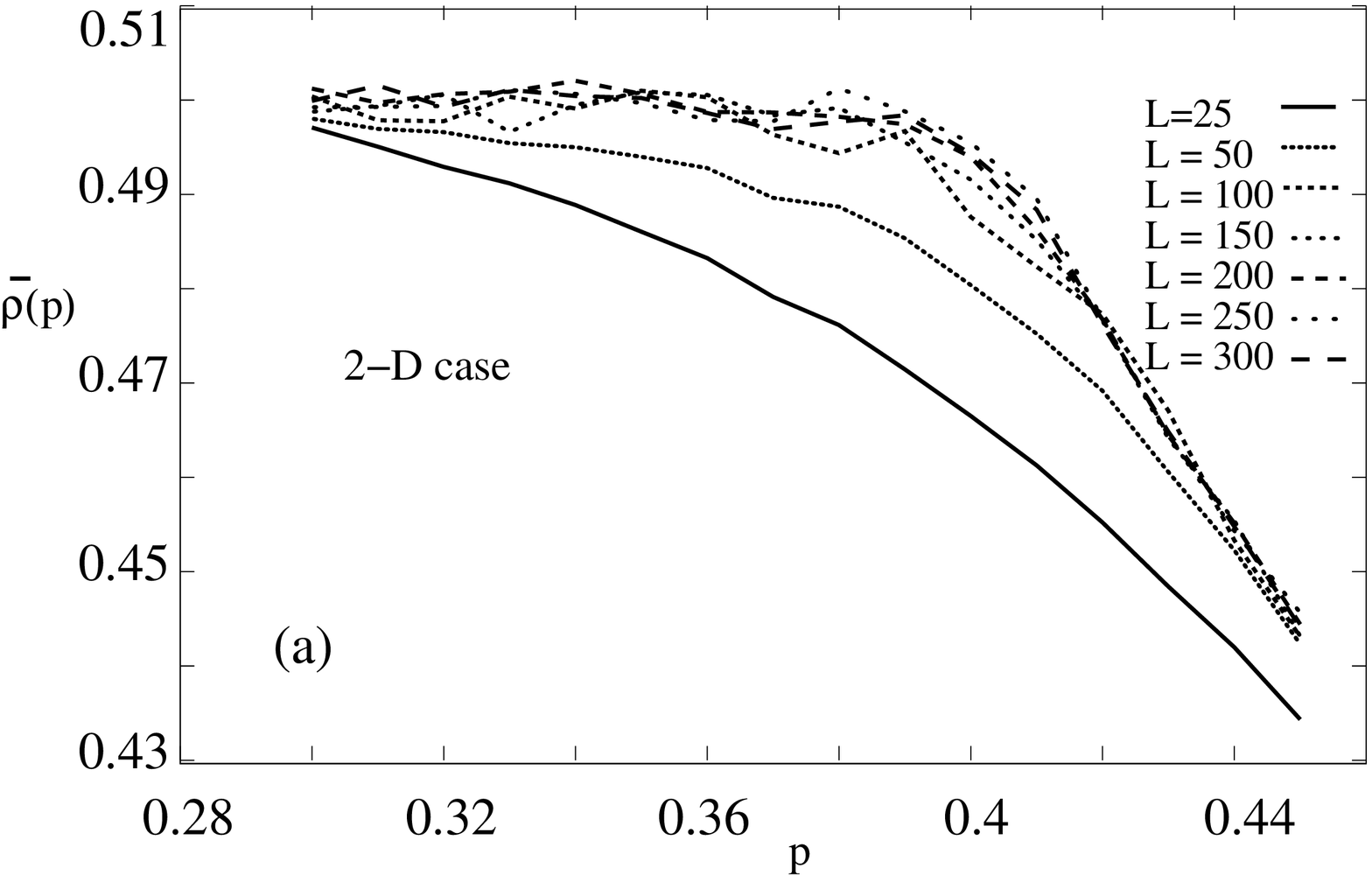}
        \includegraphics[ width=6cm]{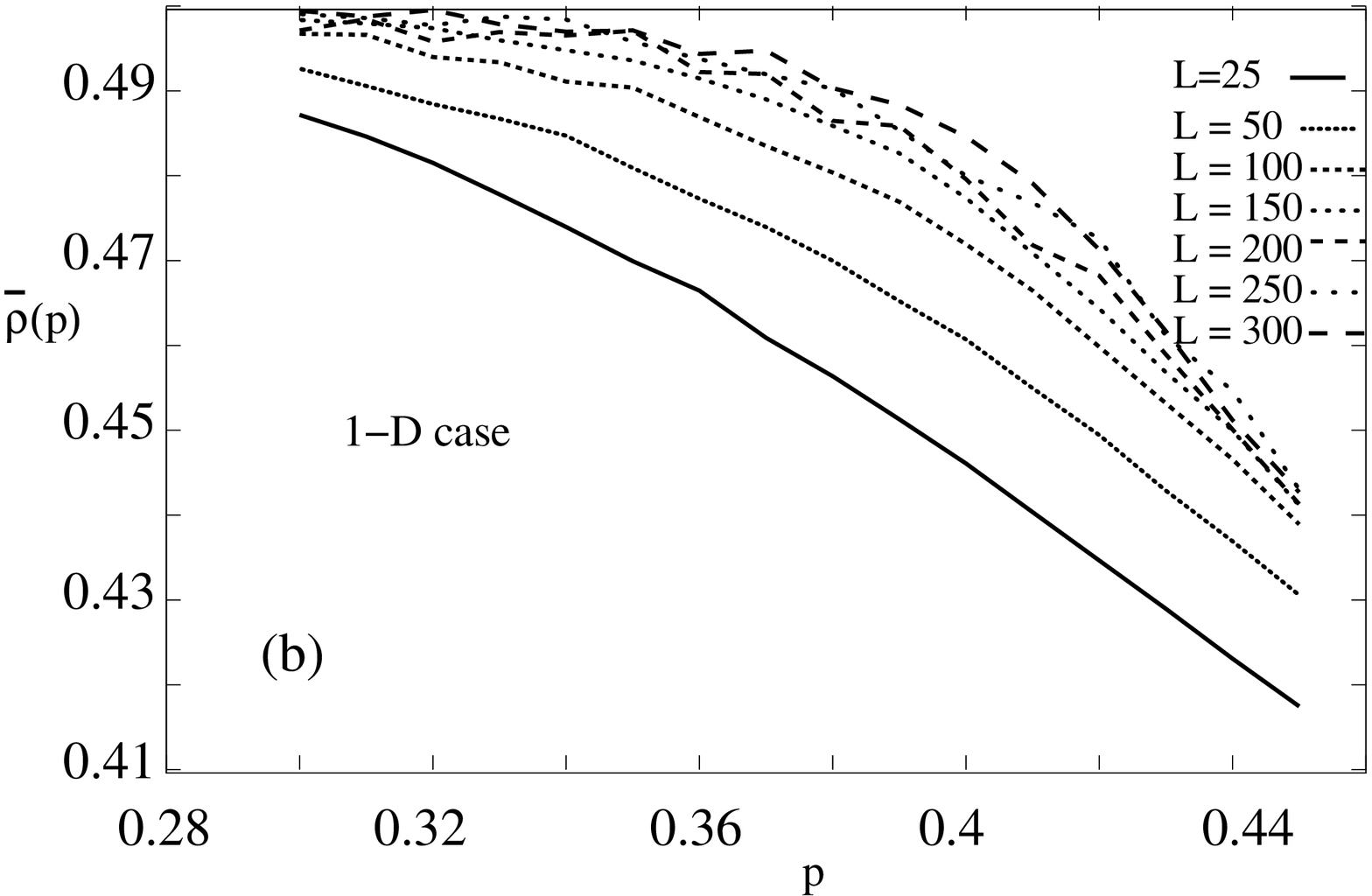}
        \includegraphics[width=6cm]{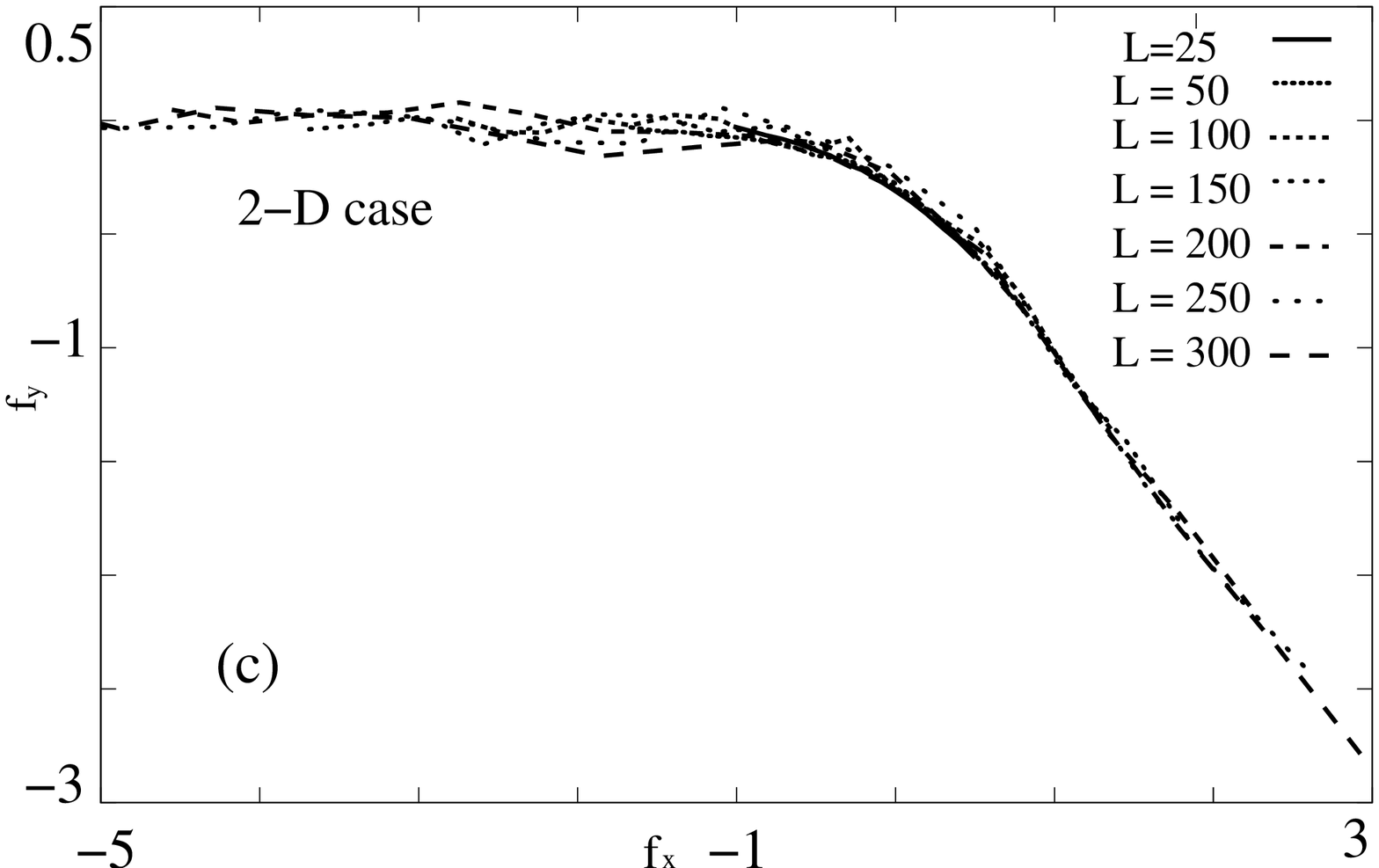}
        \includegraphics[width=6cm]{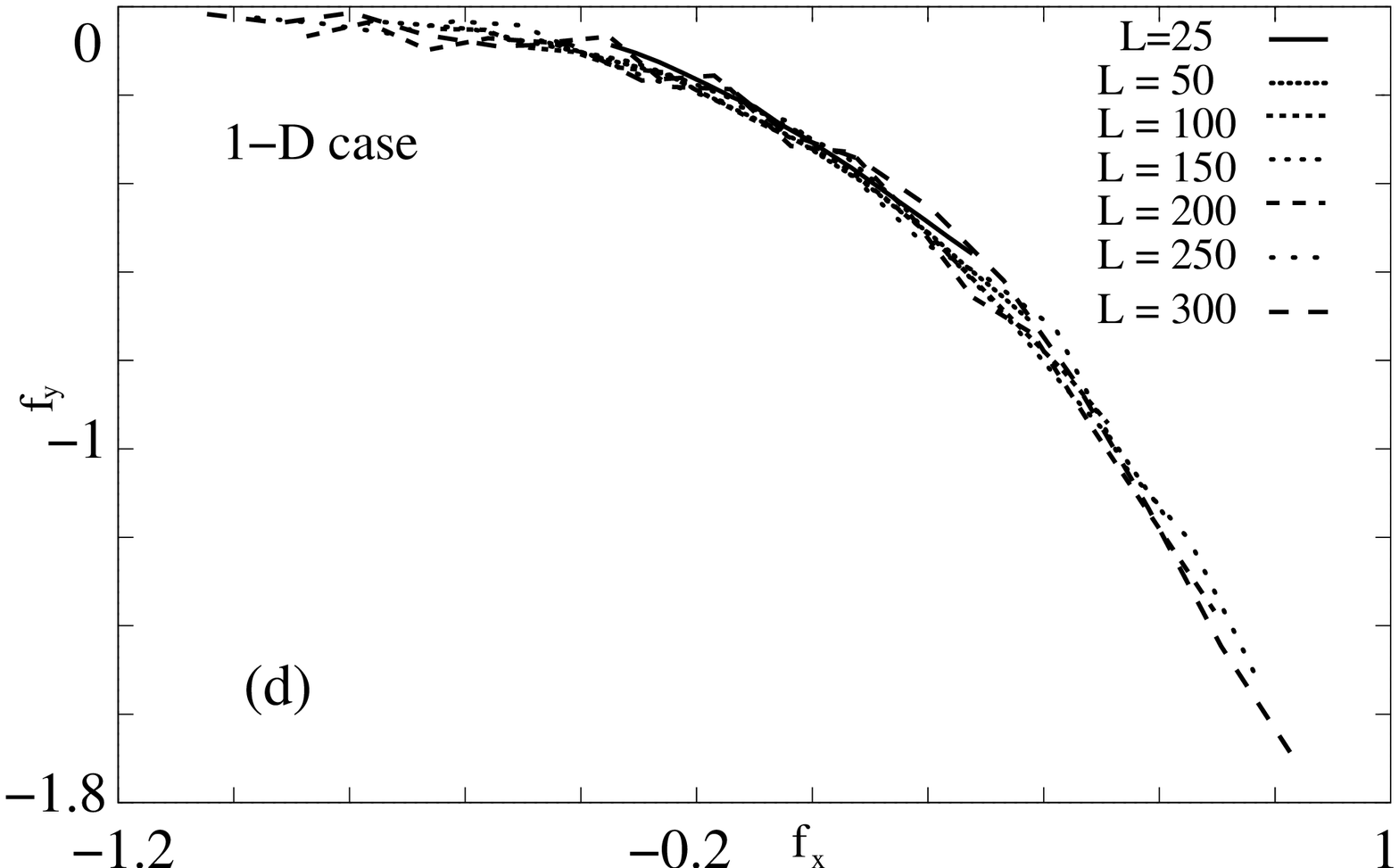}

%     \begin{tabular}{cc}
%        \includegraphics[height=3.2cm, width=4.2cm]{jt-point-2d.eps}&
%        \includegraphics[height=3.2cm, width=4.2cm]{jt-point-1d.eps}\\
%        \includegraphics[height=3.2cm, width=4.1cm]{collapse-2-D.eps}&
%        \includegraphics[height=3.2cm, width=4.1cm]{collapse-1-D.eps}
%     \end{tabular}
    \caption{Finite size effects near the transition for the 2-D case (a)
      and  for 1-D (b).  Data collapse is shown in (c) and (d), with $f_x = (p-p_c)L^{1/\nu}$ and $f_y = (\rho-\rho_c)L^{-\mu}$.}
    \label{fig:4}
  \end{figure}
%\end{widetext}
}%

%-------------------end fig 4-----------------------------------

%--------------------------fig 5------------------------------
\newcommand{\figfive}{%
  \begin{figure}
    \centering
    \begin{tabular}{cc}
       \includegraphics[height=3.1cm, width=4.2cm]{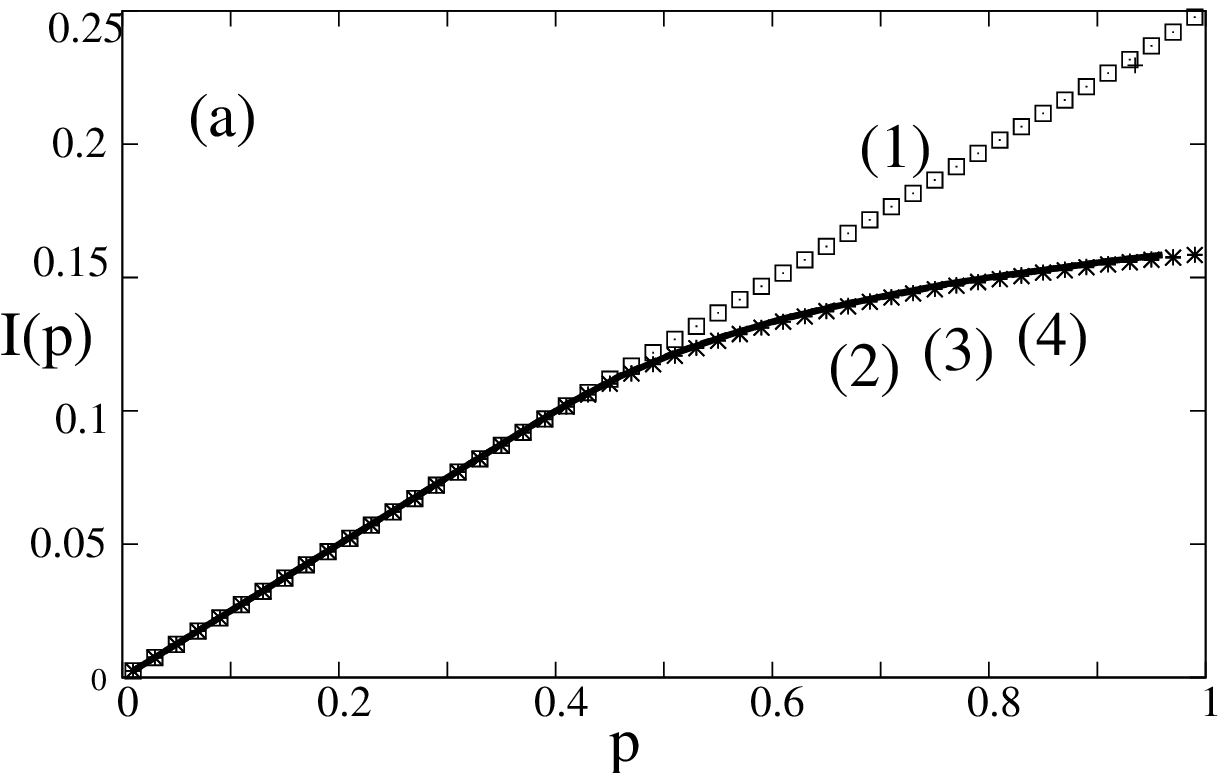}&
       \includegraphics[height=3.1cm, width=4.2cm]{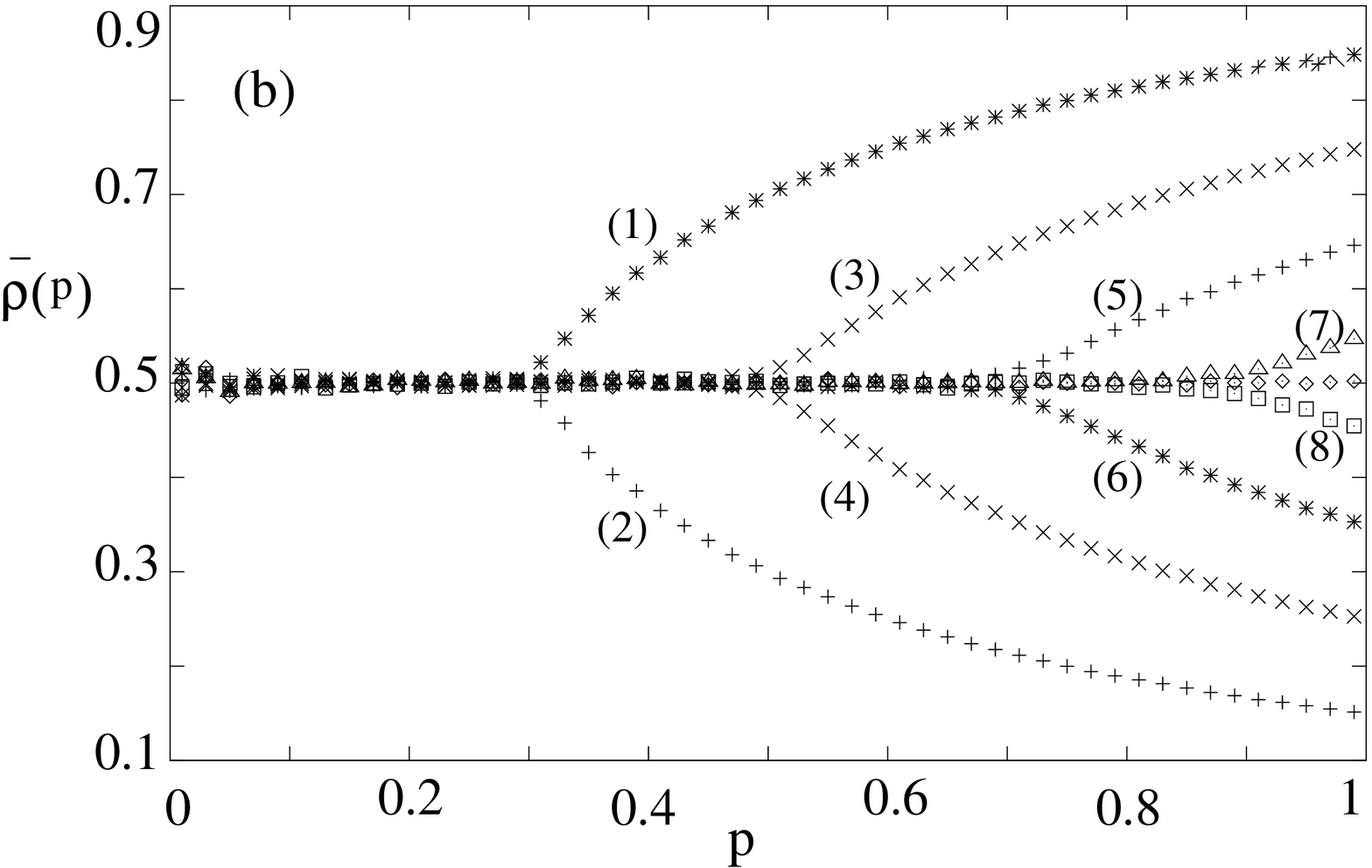}
    \end{tabular}
    \caption{(a) Current $I(p)$ vs. $p$  for 2-D case.
      Curve marked (1) is for $\alpha = 0.8$, $\beta = 0.8$, curves
      marked (2) and (3) are for 
      $\alpha = 0.2$, $\beta = 0.8$ and $\alpha = 0.8$, $\beta = 0.2$
      respectively, and (4) for $\alpha = 0.2$, $\beta = 0.2$.
      Currents are rescaled to match the bulk value $\rho(1-\rho)$
      at 
      $p=1.0$, to correct for finite size effects. Similar result
      holds for currents in  
      1-D case also. The solid line is 
      $I(p)=p\rho (1-\rho)$ for the parameters of curves (2) and
      (3). In (b) the dependence of $p_c$ 
      on $\alpha$ and $\beta$ is shown.  For all the curves,
      $\alpha+\beta=1$.  The curves are: (1) $\alpha = 0.85$, 
      %$\beta = 0.15$; for 
      (2) $\alpha = 0.15$,% $\beta = 0.85$; for
      (3) $\alpha = 0.75$, %$\beta = 0.25$; for
      (4) $\alpha = 0.25$, %$\beta= 0.75$; for
      (5) $\alpha = 0.65$, %$\beta = 0.35$; for
      (6) $\alpha = 0.35$, %$\beta = 0.65$; for
      (7) $\alpha = 0.55$, %$\beta = 0.45$; for
      (8) $\alpha = 0.45$, %$\beta = 0.55$.
      The main observation is that $p_c=2\alpha$ for downward curves
      ($\alpha$-phase region of Fig.  \ref{fig:1}(b)) and $p_c
      =2\beta$ for the up going curves ($\gamma$-phase region of Fig.
      \ref{fig:1}(b)).}
    \label{fig:5}
  \end{figure}
}%

%----------------------------end fig 5------------------------------

%----------------------fig 6------------------------------

%\newcommand{\figsix}{%
%  \begin{figure}
%          \includegraphics[height=5cm, width=6.5cm]{bound-p-1-2D.eps}
%           \caption{The importance of bulk and boundary; the density
%      transition is missing when we consider only boundary sites
%      (finite $p$) and make $p=1$ in all interior sites. Here $\alpha = 0.2$ and $\beta = 0.8$.}
%    \label{fig:6}
%  \end{figure}
%}%

%-------------------end fig 6-----------------------------------

%----------------------fig 6------------------------------

\newcommand{\figsix}{%
  \begin{figure}
    \centering
    \begin{tabular}{cc}
	\includegraphics[height=3.5cm, width=4.2cm]{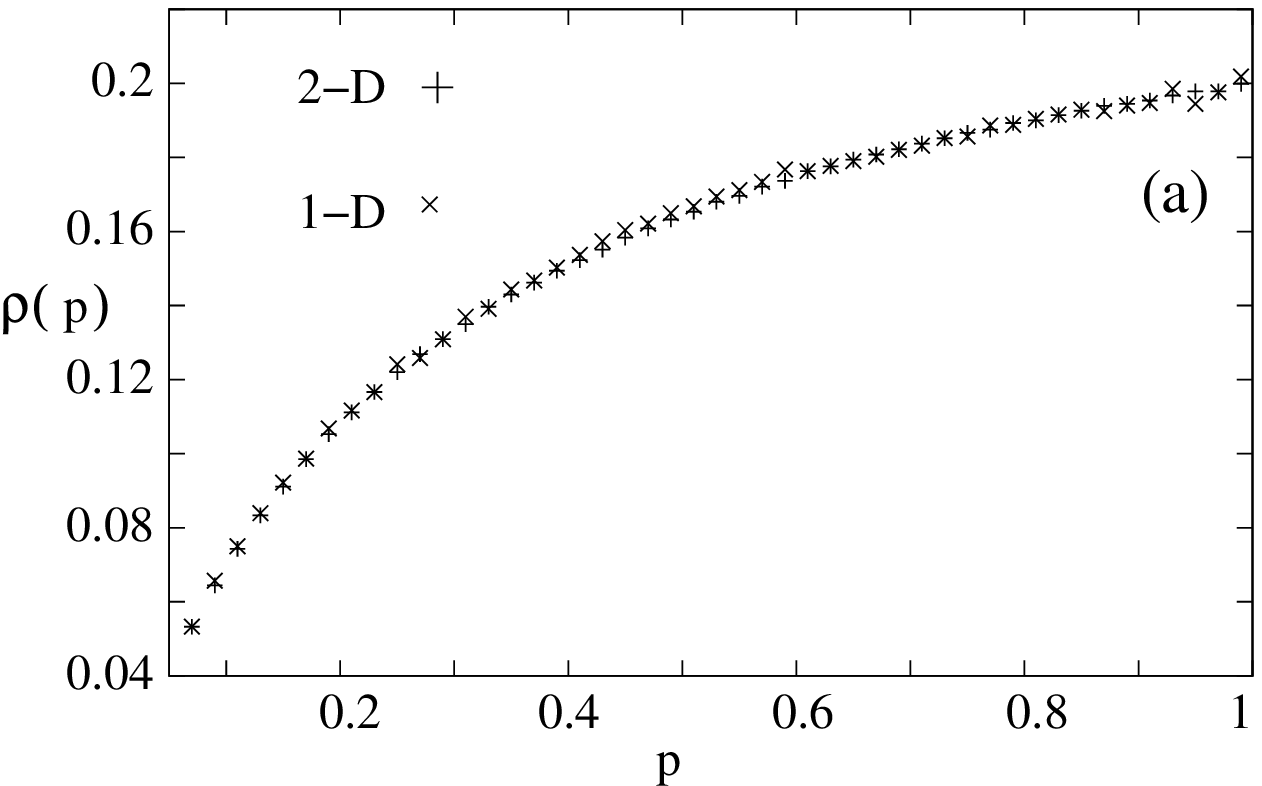}
       \includegraphics[height=3.5cm, width=4.2cm]{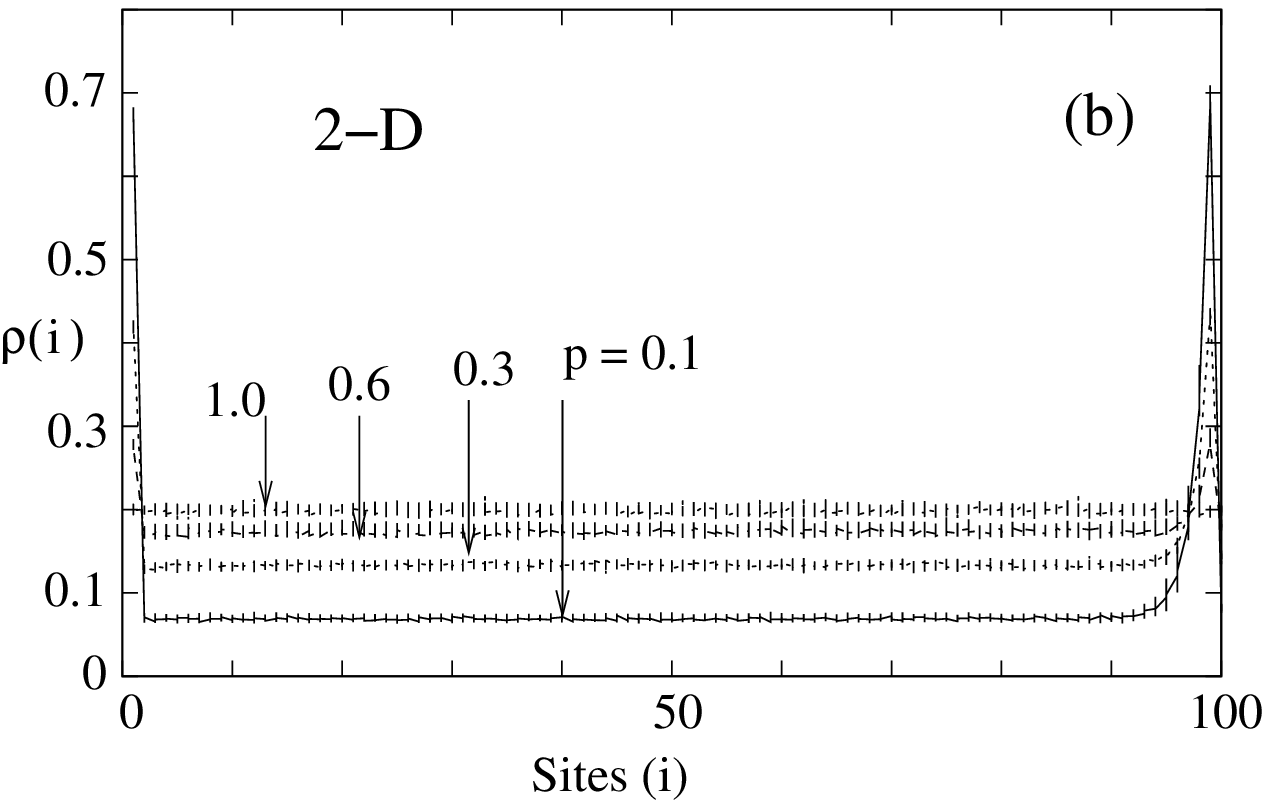}
    \end{tabular}
    \caption{The importance of bulk and boundary (a) the density
      transition is missing when we consider only boundary sites
      (finite $p$) and make $p=1$ in all interior sites. (b) spatial density profiles in 2-D case for various values of $p$. For $\alpha,\gamma=0.2$. Similar behaviour is observed in equivalent 1-D case.}
    \label{fig:6}
  \end{figure}
}%

%-------------------end fig 6-----------------------------------
%------------------------------------------------------------------------------

\title{Transverse diffusion induced phase transition in asymmetric
  exclusion process on a surface}

\author{Navinder Singh and Somendra M Bhattacharjee}

\email{navinder@iopb.res.in, somen@iopb.res.in}

\affiliation{Institute of Physics, Bhubaneswar-751005, India} 

\begin{abstract}
  We extend one dimensional asymmetric simple exclusion process (ASEP)
  to a surface and show that the effect of transverse diffusion is to
  induce a continuous phase transition from a constant density phase
  to a maximal current phase as the forward transition probability $p$
  is tuned. The signature of the nonequilibrium transition is in the
  finite size effects near it.  The results are compared with similar
  couplings operative only at the boundary.  It is argued that the nature 
  of the phases can be interpreted in terms of the modifications of
  boundary layers.
\end{abstract}  

%PACS numbers: 05.40.-a, 02.50.Ey, 64.60.-i,89.75.-k\newline\\


\maketitle

\section{Introduction}\label{sec:1}
History has shown us that the study of model systems or toy models of
real physical systems is the first step towards a deeper understanding
of working of real physical systems\cite{peierls}. In this spirit,
asymmetric simple exclusion process (ASEP) is a prototypical model of
non-equilibrium statistical mechanics that deals with systems with
currents flowing through them. Such systems are, in general externally
driven, for example, in living cells, motor proteins, traffic flows,
driven diffusive systems, transport in condensed matter and mesoscopic
systems etc\cite{stinch,ligett}.

\figone

ASEP is comprised of particles moving in a particular direction with
the constraint of no two particles at the same site at the same time,
called simple exclusion. A particle can hop if the next site is empty.
Particles are fed at one end, say $i=1$ at a rate $\alpha$ and
withdrawn at $i=m (m\to\infty)$ at a rate $\beta=1-\gamma$ so that
there is a current through the track. See Fig. \ref{fig:1}(a).  The main
interest in ASEP has been in the steady state properties, especially
the nonequilibrium phase diagrams and the stability of phases as the
external parameters or drives are changed. The phase diagrams in
several cases are known both for conserved and nonconserved
cases\cite{evans,derrida} and an intuitive deconfinement of boundary
layer approach provides a physical picture of the phase
transitions\cite{sutapa,smvm,sutapa1}. Several variants of ASEP have
also been studied\cite{ali,jaya,sutapa2}.

For 1-D ASEP chains, the phase diagram for the case with conservation
in the bulk is shown in Fig.\ref{fig:1}(b).  For large length, the
phases are characterized by the density $\rho(x), x\in [0,1]$. The
external drives at the boundaries maintain a density $\rho=\alpha$ at
$x=0$ and $\rho=\gamma\equiv 1-\beta$ at $x=1$, and determine the fate
of the bulk phase.  Unlike equilibrium situations, the information of
the bulk phases and phase transitions are contained in the boundary
behaviour. (This can be termed a ``holographic principle").  In the
$\alpha$-phase of Fig.\ref{fig:1}(b), the bulk density is
$\rho(x)=\alpha$ with a thin boundary layer maintaining the density at
the other end.  Similarly, in the $\gamma$-phase, $\rho(x)=\gamma$ in
the bulk with a boundary layer at the $x=0$ end.  There is a maximal
current phase with $\rho(x)=0.5$ for $\alpha\geq 0.5,\gamma\leq 0.5$
with boundary layers on each side protecting the bulk.  In all these
cases, the boundary layers are attached to the edges.  On the first
order phase boundary between the $\alpha$- and the $\gamma$-phase, the
density profile is $\rho(x)=\alpha+(\gamma -\alpha) x$ without any
boundary layer.  In case of a non-conservation in the bulk, this phase
boundary gets replaced by a shock phase with localized shocks on the
track\cite{evans}.  This additional shock phase can be understood as a
deconfinement transition of the shock from the boundary\cite{sutapa}.

Here, we consider a collection of such one dimensional ASEP chains
diffusively coupled to form a two dimensional ASEP (2-D ASEP).  The
transverse diffusion does not lead to any current in the extra
dimension but affects the bulk and boundary in the preferred forward
direction.  An arbitrary chain may seem to have nonconservation
through the leakage to or from the neighbouring chains but there is an
overall bulk conservation on the lattice.  We show here from
simulations the existence of the maximal current phase with $\rho=0.5$
for high transverse coupling over a wider range of $\alpha$ and
$\gamma$ with a phase transition to the
conventional phase at a critical coupling. The phase transition
behaviour in this situation can be analyzed through the changes in the
boundary layers.  To do so, we also consider a few variants of the
model both in one and two dimensions.

\section{Model}\label{sec:2}
Consider a modified asymmetric exclusion process (ASEP) on a sheet of
$ m \times n$ sites as shown in Fig. \ref{fig:1}(a) with forward
particle jump probability $p>0$ and the transverse (perpendicular to the
forward direction) probability $q$ (with the constraint $p+2q = 1$)
provided the neighboring sites are empty.  Here $q$ is a measure of
the transverse coupling of the chains.  For $q=0$, we get back
independent 1-D ASEP chains.  On the left boundary $i = 1, 1\le j \le
n$ particles are injected at a rate $\alpha$ and on the right boundary
$(i = m, 1\le j \le n)$ particles are withdrawn at a rate $\beta$. The
sheet is folded in a cylindrical geometry to impose periodic boundary
conditions in the transverse direction i.e., sites $(i,j=1)$ are
identified with sites $(i,j=n+1)$. Thus in the steady state situation
we have a net particle current in the forward direction only, and no
particle current in the transverse direction, because the
probabilities of up- and down-hops are the same.

\figtwo

\subsection{Mean field analysis}\label{sec:21}
The occupation number at site $(i,j)$ is $\tau_{i,j} = 0$ or $1$ depending
upon whether the site is empty or occupied. The rate equation
governing the average particle density distribution $\rho(x,y) \equiv
\langle \tau_{i,j}\rangle$ (where the average is over all realizations
of the process) in the bulk is:
\begin{subequations}
\begin{equation}
  \frac{d\la \tau_{i,j} \ra}{dt} =
  p[\la(1-\tau_{i,j})\tau_{i-1,j}\ra-\la(1-\tau_{i+1,j})\tau_{i,j}\ra]
  +  J_{ij}^{\rm{T}},\phantom{x}\label{eq:3} 
\end{equation}
where the transverse part is 
\begin{eqnarray}
  \label{eq:1}
J_{ij}^{\rm{T}}&=&q[-\la(1-\tau_{i,j+1})\tau_{i,j}\ra-\la(1-\tau_{i,j-1})\tau_{i,j}\ra\nonumber\\
   &&+\la(1-\tau_{i,j})\tau_{i,j+1}\ra+\la(1-\tau_{i,j})\tau_{i,j-1}\ra].  
\end{eqnarray}
The rate equation for the two boundaries are
\begin{eqnarray}
\frac{d\la \tau_{1,j} \ra}{dt} &=& +\alpha\la1-\tau_{1,j}\ra-
p\la(1-\tau_{2,j})\tau_{1,j}\ra +   J_{1j}^{\rm{T}},\phantom{xxxx}\label{eq:6}\\
\frac{d\la \tau_{m,j} \ra}{dt} &=& -\beta\la\tau_{m,j}\ra+
p\la(1-\tau_{m,j})\tau_{m-1,j}\ra +  J_{mj}^{\rm{T}}.\label{eq:7}\hfill
\end{eqnarray}
\end{subequations}
It is interesting to note an invariance in the above equations known
as the particle-hole symmetry. It implies that if we change $\alpha$
to $1-\beta$ and $\beta$ to $1-\alpha$ with $\tau$ changed to
$1-\tau$, the equations of the process remains invariant.

\figthree

In a mean-field independent-site approximation, one sees that
$\rho=$constant is a solution of the bulk equation in the steady
state.  The phase of the system is then determined by the boundary
conditions.  It transpires that a constant density cannot satisfy in
general both the boundary conditions.  This importance of the
boundary, {\it i.e.}, the choice of one, both or none of the boundary
conditions, is at the heart of the phase transitions.  One can in
addition do a stability analysis to see that a constant bulk density
is indeed a stable solution\cite{Molera}.  With the Boltzmann
approximation (neglecting nearest neighbour correlations) i.e.,
$\la\tau_{i,j}(1-\tau_{i\pm1,j\pm1})\ra \equiv
\la\tau_{i,j}\ra(1-\la\tau_{i\pm1,j\pm1}\ra),$, we take $
\la\tau_{i,j}\ra = \rho_0 + \delta \rho(i,j,t)$, with $\delta\rho$ a
small perturbation. In terms of the Fourier modes,
\begin{equation}
  \label{eq:4}
\delta{\rho({\bf{k}},t)} = \sum_{x,y} e^{-i
  k_x x - i k_y y} \delta\rho(x,y,t),\qquad  k_{x(y)} = \frac{2\pi
  q_{x(y)}}{L},  
\end{equation}
where $L$ is the dimension of the lattice, and $\bf{k}$
denotes $\{k_x,k_y\}$, Eq. \ref{eq:3} can be written as 
\begin{equation}
 \frac{d\delta{\rho({\bf{k}},t)}}{dt} = \Omega(p,q,{\bf{k}}) \delta{\rho({\bf{k}},t)},
\end{equation}
with 
\begin{equation}
  \label{eq:5}
  \Omega(p,q,{\bf{k}}) = i p \sin k_x (2\rho_0-1) + p \cos k_x + (1-p)\cos k_y -1.
\end{equation}
Since $p \cos k_x + (1-p)\cos k_y -1 < 0$, the negativity of the real
part of $\Omega$ insures decaying perturbations and stability.  This
linear stability analysis, though useful in the context of traffic
jams in similar two-dimensional models\cite{Molera}, is not enough for
ASEP.

\subsection{Simulation}\label{sec:22}
To simulate the process for any $p$ we  use a random sequential
update scheme. Starting from a random distribution, we allow the
system to reach a steady state. From the simulations we study the
spatial density distribution  and currents for various
values of $p$ and for various sizes of the lattice. 

To analyze in detail the density dependence on $p$, let us define the
average bulk density $\bar{\rho}(p)$ (for given $\alpha$ and $\beta$),
\begin{equation}
\bar{\rho}(p) = \frac{1}{N}\sum_{cyc = 1}^N\frac{1}{n^\prime} \sum_{i,j\subset A_{center}} \tau_{i,j}(p).
\end{equation}
The averaging in Eq.(5) is done on a strip ($A_{center}$) at the
center of the cylinder i.e.,$(m/2-4 < i < m/2+4, 1 < j < n)$), and 
$n^\prime$ ($=9\times n$ in our case) is the number of sites in the
central strip $A_{center}$.   $N$ in the above expression is the total
number of cycles of the simulation $(\sim 10^6)$ used for averaging.

\figfour

\section{Results}\label{sec:3}
For given $\alpha$ and $\beta$, the steady state profiles are of two
types as shown in Fig. \ref{fig:2}(a) for several values of $p$.  We see
that for small $p$ the bulk reaches half-filling and changes over to a
boundary dependent density for larger $p$.  In Fig. \ref{fig:3},
$\bar{\rho}$ is plotted as a function of $p$ for various values of
$\alpha$ and $\beta$. The behaviour shown in Fig.  \ref{fig:2}(a) is
evident here.  For the case $\alpha > 0.5$, $\beta < 0.5$, the
behaviour is complementary to the case $\alpha < 0.5$, $\beta > 0.5$.
However,  there is no such transition for $\alpha > 0.5$, and
$\beta > 0.5$.
The critical $p_c$ also depends upon the values of $\alpha$ and
$\beta$ as shown in the phase diagram Fig.  \ref{fig:1}(b). 
We have
studied an equivalent 1-D model, because the transverse periodic
boundary conditions in 2-D has some similarity with 1-D. To mimic the
behaviour we modify the 1-D ASEP so that a particle jumps to the next
empty site not with probability $1$ but with probability $p$, i.e.,
particle waits with probability $1-p$. The average bulk density in
this 1-D case also shows behaviour similar to the 2-D case and is  shown in
Fig. \ref{fig:3}(b).  One sees the phase transition with $p$.  Thus,
we have the following three main observations:
\begin{enumerate}
\item $\rho(i,j) = 0.5$ for all $p$ less than $p_c$, for all $\alpha$
  and $\beta$.
\item In the regime $p>p_c$, mean-field continuum approximation is
  valid and phase diagram resembles the  1-D phase diagram.
\item In the shaded region marked $\alpha$-phase in Fig.
  \ref{fig:1}(b) $p_c = 2\alpha$ and in the $\gamma$-phase region
  $p_c=2\beta$.
\end{enumerate}
These results can be explained by examining the boundary densities.
If we do a mean-field approximation in the steady-state situation of
Eq.  \ref{eq:6}, then a homogeneous density would give
$(\alpha-p\rho)(1-\rho) = 0$ or $\rho=\alpha/p$ for the left boundary.
The bulk current is expected to be $I(p)=p\rho(1-\rho)$ (see below) as
shown in Fig. \ref{fig:5}(a).  For ASEP, the bulk satisfies the left
boundary condition only in the $\alpha$-phase which requires the
boundary density to be less than or equal to $0.5$.  Therefore a
maximal current phase is expected if $\alpha/p>0.5$, {\it i.e.}, $p_c
= 2\alpha$.  The left boundary layer then develops
(Fig.\ref{fig:2}(a)) for $p < p_c$.  The density variation in the
boundary layer vitiates the simple argument because the density
gradient dependent diffusive part of the boundary current needs to be
taken into account.  The net boundary density is obtained by the
balance of the input and the outflow consisting of the hopping and the
diffusive parts.  Similar argument holds in the $\gamma$-phase region
for $\beta$ and $p_c$.  For the $\gamma$-phase, the right density is
$\rho=[\gamma-(1-p)]/p$ if there is no boundary layer.  The bulk
density is controlled by this boundary value (rather than the
withdrawal rate) so that it also takes the same value as the boundary.
These observations are supported by Fig.  \ref{fig:5}(a,b).

\figfive

It is known for ASEP, that on the first order phase boundary
separating the $\alpha$- and the $\gamma$-phases, there are shocks
that diffuse slowly on the track vanishing or getting created at the
boundaries only.  Same thing happens here also on the phase boundary
which is still set by $\alpha=1-\gamma$.  Because of slow diffusion of
the shock, the measured density in the central patch could be either
that of the $\alpha$-phase or of the $\gamma$-phase.  This is shown in
Figs. \ref{fig:3} (a) and (b)).  The density remains constant for
$p<p_c$, but after this ($p>p_c$) average density shows an erratic
behaviour, fluctuating wildly\cite{evans}.  The special point where
the three phase boundaries meet is now at
$\alpha=p/2,\gamma=1-(p/2)$ in the $\alpha,\gamma$ plane.

The above meanfield results seem to suggest a singularity in the
density as a function of $p$, because $\bar{\rho}=0.5$ for $p<p_c$ but
$\bar\rho = \alpha/p$ for $p>p_c$. Such a singularity is expected only
in the long chain limit (infinitely long system) and not in finite
systems.  Fig. \ref{fig:4}(a,b) shows a strong size dependence near
$p_c$.  For equilibrium phase transitions, singularities are rounded
off by finite size when the size of the system is comparable to the
characteristic length scale for the transition.  The finite size
behaviour, especially the size dependence, then follows a finite size
scaling form.  In that spirit, let us make a finite size scaling
ansatz for this nonequilibrium case as
\begin{equation}
  \rho - \rho_c \sim L^{-\mu} f([p-p_c] L^{1/\nu}), 
\end{equation}
with,
\begin{equation}
  \label{eq:2}
\rho - \rho_c \sim |p-p_c|^{\mu\nu}  \quad {\rm for\ } L\to\infty
\end{equation}
where $\rho_c = 0.5$ is the constant density for $p<p_c$, $L$ is the
linear dimension of the system (lattice or chain),and,  $\mu$, and $\nu$ are
scaling indices, then, to recover the meanfield results, we need to have
$\mu=\nu^{-1}$.  We have used the Bhattacharjee-Seno method for
data-collapse\cite{collap}.  In Fig. \ref{fig:4}(c) the data collapse 
scaling is shown for 2-D for which we get $p_c = 0.40\pm 0.008, ~~\mu =
0.69\pm 0.07,~~\nu^{-1} = 0.72\pm 0.03$.  For  the 1-D case (Fig.
\ref{fig:4}(d)), we have $p_c = 0.401\pm 0.006, ~~\mu = 0.46\pm
0.02,~~\nu^{-1} = 0.44\pm 0.06$.  These are consistent with the
prediction of $\mu\nu=1$.  The characteristic length scale seems to
diverge as $\xi\sim \mid p -p_c\mid^{-\nu}$ which is set by the width
of the boundary layer.  Meanfield analysis is not fine enough to get
this length properly.

%\figfive

Since the current is a measure of jumps from occupied sites to nearest
vacant site in the forward direction, the probability of site
occupation is $\rho$, the probability of vacancy of the next site is
$1-\rho$, and jump probability in the forward direction is $p$, thus,
the net current in the forward direction is $I(p)=p\rho (1-\rho)$.
Consequently, $I(p)=p/4$ for $p< p_c$, while $I(p)=\alpha(1-
\alpha/p)$ for $p>p_c$, joining continuously at $p=p_c$ with a slope
discontinuity.  Fig. \ref{fig:5}(a) shows the overall agreement of the
measured current and this general form of the current when the
correnponding $\bar{\rho}$ obtained from the simulation is used.
However, finite size rounding masks the expected singularity at
$p=p_c$ in this current plot.
%\figeight
\figsix

In order to show that the above results, though boundary driven, are not
a consequence of local perturbations at the boundary, we considered a
variant of the model where the transverse coupling is only at the two
ends.  We have put $p=1$ in the all the bulk sites i.e., for sites $2
\le i \le n-1 $ and kept finite $p$ for the first and the last site,
{\it i.e.}, jumps from first site to second and $n-1$th site to $n$th
happen with finite $p$.  We see that the system self-organizes to a state
with new boundaries that control the bulk density.  The actual
drives (the injection and withdrawal rates) passively help in creating
the relevant boundary conditions.  In 
particular, we observe that the transition induced by $p$ for the bulk
case is no longer present.
The behavior of average $\rho$ with $p$ is shown in Fig. \ref{fig:6} (a)
and the corresponding density profiles are shown in Fig. \ref{fig:6}(b) (similar profile has benn observed in 1-D case also). The 
behaviour of average $\rho$ with very small $p < 0.05$  shows a long living transient state, due to the very small forward motion. These observations indicate that the transition is due to a
co-operative phenomenon, where bulk and boundary play their role
co-operatively and inter-dependent way.

\section{summary}

In conclusion, the continuous transition from the injection rate
dominated phase to the maximal current phase has been observed as a
function of forward transition probability $p$ in a two dimensional
ASEP (diffusively coupled chains).  The transition shows finite size
effects, reminiscent of equilibrium phase transitions, and finite size
scaling predicts exponents which are consistent with the mean field
theory predictions.  The bottleneck created at the boundary by the
transverse coupling changes the effective particle densities at the
two boundaries and the ensuing phase diagram can then be mapped out
from the 1-d phase diagrams with $p=1$, with the multicritical point shifting to $(\alpha=\frac{p}{2}, \gamma = 1-\frac{p}{2})$.  However no such transition
can be induced if artificial bottlenecks are created at the boundaries
only.  In such situations, the particles organize themselves to form a new or
effective boundary density which then  as per the holographic principle
 fixes the bulk density.  This reiterates that the
nonequilibrium transitions observed are cooperative but boundary
driven and the boundary layers contain the information about the bulk.

%---------------------------------


\begin{thebibliography}{100}

\bibitem{peierls}Sir Rudolf Peierls, {\it Model-making in physics}, Contemp. Phys. {\bf 21} 3 (1980).

%

\bibitem{stinch}R. B. Stinchcombe, Adv. Phys. {\bf 50} 431 (2001).

%

\bibitem{ligett} T. Ligett, {\it Interacting Particle Systems: 

Contact, Voter and Exclusion Processes} (Springer-Verlag, Berlin, 1999)

%

\bibitem{evans}M. R. Evans, R. Juhasz and L. Santen, Phys. Rev. E {\bf
    68} 026117 (2003); A. Parmeggiani, T. Franosch and E. Frey, Phys.
  Rev. E 70, 046101 (2004).

%

\bibitem{derrida}B. Derrida, Phys. Rep. {\bf 301} 65 (1998).

%

\bibitem{sutapa}S. Mukherji and S. M. Bhattacharjee, J. Phys. A {\bf 38} L285 (2005).

%

\bibitem{smvm}S. Mukherji and V. Mishra, Phys. Rev. E {\bf 74} 011116 (2006).

%

\bibitem{sutapa1}S. Mukherji, Physica A: Stat Mech and its App, {\bf 384}, 83 (2007).


\bibitem{ali} M. Alimohammadi, V. Karimipour, M. Khorrami, J. Stat. Phys. {\bf 97} 373 (1999).

%

\bibitem{jaya} Jaya Maji and S. M. Bhattacharjee, Euro. Phys. Lett. {\bf 81} 30005 (2008).

%

\bibitem{sutapa2} Sutapa Mukherji, Phys. Rev. E {\bf 76}, 011127 (2007).

%

\bibitem{collap} S. M. Bhattacharjee and F. Seno,  J. Phys. A 34,  6375 (2001).
\bibitem{Molera}Jaun M. Molera, F. C. Martinez, J. A. Cuesta, and Ricardo Brito, Phys. Rev. E. {\bf 51} 175 (1995).

\end{thebibliography}
\end{document}